\documentclass[12pt]{article}
\usepackage{amssymb}

\usepackage{amsmath}

\begin{document}

\title{Raman scattering due to disorder-induced polaritons}
\author{Lev I. Deych$^{\S }$, M. Erementchouk$^{\S \S }$, and V.A. Ignatchenko$^{\S
\S }$ \\
$^{\S }$Department of Physics, Seton Hall University, South Orange, NJ 07079%
\\
$^{\S \S }$ Kirensky Institute of Physics, Krasnoyarsk, 660036, Russia\\
Department of Physics}
\maketitle

\begin{abstract}
The selection rules for dipole and Raman activity can be relaxed due to
local distortion of a crystalline structure. In this situation a
dipole-inactive mode can become simultaneously active in Raman scattering
and in dipole interaction with the electromagnetic field. The later
interaction results in disorder-induced polaritons, which could be observed
in first-order Raman spectra. We calculate scattering cross-section in the
case of a material with a diamond-like average structure, and show that
there exists a strong possibility of observing the disorder induced
polaritons. \medskip

\noindent PACS numbers: 78.30-j, 63.50+x, 63.20-e.
\end{abstract}

\section{\protect\bigskip Introduction}

The effects of a resonance linear interaction between excitations of
different nature are well known and their importance for properties of
crystalline materials has been appreciated for a long time. One could
mention, for instance, such phenomena as magneto-elastic resonance \cite
{magnetoelastic}, or various kinds of the polariton resonance \cite
{polariton}. The former refers to resonance interaction between spin and
elastic waves that results in mixed excitations carrying both magnetic and
elastic types of energy. The latter term is used to refer to excitations
arising as a result of interaction between electromagnetic waves and
excitations of crystals such as phonons, excitons, plasmons, and so on. In
spite of significant differences between all the cited examples, they share
one common feature. In the absence of interaction the dispersion curves of
the participating excitations cross each other at a certain point. This
crossing is actually responsible for the resonance nature of the
interaction, which lifts the degeneracy of the initial excitations at the
crossing point and gives rise to the mixed states.

In a recent paper \cite{DICR} the effect of crossing resonance has been
studied in systems with a \emph{random }interaction between participating
excitations. It was shown that even a random coupling parameter with zero
mean value can, under certain conditions, result in the strong coupling
between interacting waves, leading to a lifting of the degeneracy at the
crossing point. The mixed excitations arising in this model of \emph{%
disorder-induced crossing resonance }(DICR), however, differ significantly
from the respective excitations in the deterministic case. First of all,
excitations that are actually mixed together are coherent excitations of one
nature with nonzero average amplitude and \emph{non-coherent scattered
excitations of the other nature with average amplitude equal to zero}.
Correspondingly, one has to consider two such mixed excitations, and each of
them can result in dispersion laws split at the crossing point in two
branches . It is possible, therefore, to have three or even four different
dispersion curves describing averaged properties of the system. Recently the
model of DICR was developed further to include nonzero mean value of the
random coupling parameter \cite{nonzero}. It was shown in Ref.\cite{nonzero}
that transition between the completely disordered case with zero mean to the
nonrandom situation occurs in a nontrivial way. The idea of the DICR was
first applied to the magnetoelastic resonance in amorphous
zero-mean-magnetostriction alloys \cite{magnetoelastic_our}. In
magnetoelastic case it is possible to perform a direct observation of
elastic and magnetic susceptibilities modified due to DICR. The relationship
between maxima of the susceptibilities and solutions of the dispersion
equations for magnetoelastic resonance was studied in Ref. \cite
{susceptibilities}

In Ref.\cite{DIP} it was suggested that DICR can also occur for the
polariton resonance in disordered materials if one considers the interaction
of electromagnetic waves with phonon or exciton excitations that would be
dipole-inactive in the absence of disorder (disorder-induced polaritons).
Polariton dispersion curves in ideal crystals can be experimentally studied
by means of Raman scattering (see, for example, Ref. \cite{raman}). It was
also suggested in Ref.\cite{Maradudin} that polaritons arising due to
retardation effects in the electromagnetic interaction between local
impurity vibrations can also be observed in Raman scattering. It seems
natural, therefore, to explore whether it is possible to use the Raman
effect to observe the disorder-induced polaritons. In this paper we present
calculations of the first order Raman scattering cross-section from
forbidden polaritons arising in a simple cubic diamond-like structure due to
local structural distortions.

\section{Intensity of Raman scattering from disorder-induced polaritons.
General expression}

The polarization, $\mathbf{P}$\textbf{,} of a dielectric medium subjected to
an external electromagnetic field, $\mathbf{E}_{in}$,with frequency $\omega
_{in}$ can be expressed by means of \ a susceptibility tensor $\chi _{ij}$ as

\begin{equation}
P_{i}=\chi _{ij}E_{in}^{j}.  \label{eq.1}
\end{equation}

In the frequency region of our interest, susceptibility can be divided into
two major contributions: ion susceptibility $\chi _{ij}^{ion}$ and electron
susceptibility $\chi _{ij}^{el}$. The former part of the susceptibility
arises due to the coherent motion of ions caused by the incident
electromagnetic wave and is responsible for a modification of the wave's
velocity (real part) and for its absorption (imaginary part). The electron
contribution $\chi _{ij}^{el}$ reflects the contribution of electron
transitions in the dipole moment of the atoms constituting crystal. For
frequencies far enough from those of electron transitions and in the linear
approximation $\chi _{ij}^{el}$ can be considered as a frequency independent
constant. However, if one takes into account nonlinear corrections such as
an interaction between electrons and ionic vibrations, the electron
contribution becomes modulated by thermal ionic vibrations giving rise to
the inelastic scattering of light known as Raman scattering. As a result of
this scattering the frequency, $\omega _{in}$, and wave number $k_{in}$, of
the incident wave changes. In the case of scattering from an ideal crystal,
frequencies and wave numbers of the incident and scattered waves obey
well-known kinematic relations that reflect energy and momentum conservation
laws: 
\begin{eqnarray}
\omega _{in}-\omega _{s} &=&\pm \Omega  \label{energyconservation} \\
\mathbf{k}_{in}-\mathbf{k}_{s} &=&\pm \mathbf{K}  \notag
\end{eqnarray}
Here $\omega _{s}$ and $\mathbf{k}_{s}$ are the frequency and wave number of
the scattered field, and $\Omega $ and $\mathbf{K}$ are the frequency and
wave number of the excitations of the medium responsible for scattering. The
signs $+$ and $-$ correspond to processes of emission and absorption of the
excitations respectively giving rise to Stokes and anti-Stokes components of
the scattered field. In the presence of disorder the second of the equations
(\ref{energyconservation}) does not hold any more since disordered systems
lack translational invariance.

Intensity of the field scattered at a given direction within a certain
frequency interval is determined by the differential cross-section $%
d^{2}\sigma /dod\Omega $, where $do$ is a differential of a solid angle at a
given direction. In general the cross section can be presented as \cite
{ramanbook} 
\begin{equation}
\frac{d^{2}\sigma }{dod\Omega }=const\times
e_{i}^{s}e_{j}^{s}e_{k}^{in}e_{l}^{in}<<\delta \chi _{ik}(\Omega ,\mathbf{K}%
)\delta \chi _{jl}(\Omega ,\mathbf{K})>>  \label{crosssection}
\end{equation}
where $\mathbf{e}^{s}$ and $\mathbf{e}^{in}$ are polarization vectors of the
scattered and the incident fields respectively, and the $const$ in front of
the whole expression is a combination of such parameters as volume of the
scattering region of the sample, frequency of the incident wave, etc., which
are not significant for the purpose of our consideration. Double angular
brackets $<<\cdot \cdot \cdot >>$ denote two types of averaging:
thermodynamical average over statistical ensemble, and the average over
realizations of the random inhomogeneities of the system. Modulation of the
susceptibility $\delta \chi _{ik}(\Omega ,\mathbf{K})$ can be found if one
expands the electronic susceptibility into a power series with respect to
the amplitudes of excitations responsible for the scattering. In the case of
scattering due to polaritons, the electronic susceptibility is modulated by
both ionic displacements $W_{\sigma }$ and electric field $E$ associated
with those displacements. Therefore, $\delta \chi _{ik}(\Omega ,\mathbf{K})$
is the sum of two terms 
\begin{equation}
\delta \chi _{ik}(\Omega ,\mathbf{K})=\sum_{\sigma }\int d^{3}qW_{\sigma
}^{\ast }(\Omega ,\mathbf{q)}h_{\sigma }^{ik}(\mathbf{K}-\mathbf{q})+\int
d^{3}qE_{m}^{\ast }(\Omega ,\mathbf{q)}g^{ikm}(\mathbf{K}-\mathbf{q}),
\label{susc_expansion}
\end{equation}
where $W_{\sigma }$ represents an amplitude of $\sigma $-th normal mode of
the ionic vibrations, $h_{\sigma }^{ik}$ is the tensor of mechano-optic
coefficients, and $g^{ikm}$ represents electro-optic coefficients of the
system, wave number $\mathbf{K}$ is determined by the Eq.(\ref
{energyconservation}), and an additional integration over $\mathbf{q}$
reflects the inhomogeneous nature of the considered system and the lack of
the translational invariance. For this reason susceptibility derivatives $%
h_{\sigma }^{ik}$ and $g^{ikm}$ should be considered as random functions.
Within the phenomenological approach utilized in this paper we have to make
additional assumptions about the properties of these functions based upon
some general arguments. We begin with separating out mean parts of these
functions and represent them in the following form: 
\begin{eqnarray}
h_{\sigma }^{ik}(\mathbf{K}-\mathbf{q}) &=&<h_{\sigma }^{ik}>\delta (\mathbf{%
K}-\mathbf{q})+\tilde{h}_{\sigma }^{ik}(\mathbf{K}-\mathbf{q})
\label{h_and_g} \\
g^{ikm}(\mathbf{K}-\mathbf{q}) &=&<g^{ikm}>\delta (\mathbf{K}-\mathbf{q})+%
\tilde{g}^{ikm}(\mathbf{K}-\mathbf{q}).  \notag
\end{eqnarray}
According to this representation the scattering cross-section has two
different contributions: the coherent one, resulting from averaged parts of
the coefficients $h$ and $g$, and the non coherent contribution due to
inhomogeneous parts of the tensors. The main physical difference between
these contributions is that the former conserves momentum (this is reflected
by the presence of the $\delta $-function in the respective term of Eq.(\ref
{h_and_g})), and the latter does not. The properties of the mean parts of
the susceptibility derivatives is determined by the average symmetry of the
system. We assume the cubic average symmetry of the crystal structure that
preserves central symmetry, and that the normal vibrational mode of the
system is a triply degenerate (in the absence of interaction with
electromagnetic field) mode of even parity. (An example of such a system can
be the $\Gamma _{5}^{+}$ mode in a diamond-like structure). This mode is
dipole inactive on average and its coupling to the electromagnetic field
occurs only due to random deviations of structure from the average one. This
assumption ensures that we actually deal with the disorder-induced
polaritons, and at the same time it guarantees that $<h_{\sigma }^{ik}>$ is
not equal to zero since it is well known \cite{ramanbook} that infra-red
activity and Raman activity are mutually exclusive in the crystals with a
center of inversion. The tensor structure of $<h_{\sigma }^{ik}>$ \ in the
case of a diamond like structure can be presented in the following form \cite
{ramanbook} 
\begin{equation}
<h_{\sigma }^{ik}>d_{\sigma }^{m}=h|\epsilon _{ikm}|  \label{h_average}
\end{equation}
where $d_{\sigma }^{j}$ is a polarization vector of a given mode, and $%
|\epsilon _{ikj}|$\ is the Levi-Civita symbol.\ 

The question about properties of the electro-optical coefficient in our
model is more subtle. In the regular case of crystalline materials these
coefficients come into play only in the case of vibrational modes of odd
parity, because only in this case does there exist an electromagnetic field
associated with the vibrations. Then a third-rank tensor $g^{ikm}$ in such a
situation is equal to zero in crystals with central symmetry, therefore, the
whole issue of the properties of the electro-optical coefficients in
crystalline materials arises only in non-centro-symmetrical crystals. In our
situation of disorder-induced polaritons the mode one deals with has even
parity. Therefore, symmetry arguments do not forbid nonzero value of the
tensor $g^{ikm}$ even in materials with central-symmetry average structure.
Therefore we can assume for the tensor structure of $g^{ikm}$\ the same form
as in Eq.(\ref{h_average}).\ 

We can conclude, therefore, that the coherent contribution in the scattering
cross-section consists of two terms 
\begin{equation}
\frac{d^{2}\sigma ^{coh}}{dod\Omega }=const\times \lbrack n(\Omega
)+1](I_{W}^{coh}+I_{E}^{coh}),  \label{crosscoh}
\end{equation}
where the vibrational contribution $I_{W}^{coh}$ is 
\begin{equation}
I_{W}^{coh}=h^{2}|\epsilon _{ikm}||\epsilon
_{jlp}|e_{i}^{s}e_{j}^{s}e_{k}^{in}e_{l}^{in}\sum_{\sigma }d_{\sigma
}^{m}d_{\sigma }^{p}\mathrm{Im}\left[ \eta _{\sigma }^{W}(\Omega ,\mathbf{K)}%
\right] \mathbf{,}  \label{Wcoh}
\end{equation}

and the electromagnetic contribution $I_{E}^{coh}$ is given by the similar
expression 
\begin{equation}
I_{E}^{coh}=g^{2}|\epsilon _{ikm}||\epsilon
_{jlp}|e_{i}^{s}e_{j}^{s}e_{k}^{in}e_{l}^{in}\mathrm{Im}\left[ \eta
_{mp}^{E}(\Omega ,\mathbf{K)}\right] .  \label{Ecoh}
\end{equation}
Writing down these expressions we make use of the fluctuation-dissipation
theorem and express the vibrational and electric field correlation functions
in terms of thermodynamical Bose-factor $n(\Omega )+1$, and the respective
vibrational and electromagnetic linear-response functions $\eta _{\sigma
}^{W}(\Omega ,\mathbf{K)}$ and $\eta _{ij}^{E}(\Omega ,\mathbf{K)}$ \cite
{Loudon}.

In order to describe non-coherent contributions in the cross-section we have
to make assumptions regarding correlation properties of the susceptibility
derivatives $h_{\sigma }^{ik}$ and $g^{ikm}$. It is natural to assume that
their fluctuations are both homogeneous and isotropic. These assumptions
allow one to introduce the spectral densities (spatial Fourier-transforms)
of the corresponding correlation functions and present them in the following
form 
\begin{eqnarray*}
&<&h_{\sigma }^{ik}(\mathbf{k}_{1}\mathbf{)(}h_{\sigma }^{jl}(\mathbf{k}_{2}%
\mathbf{))}^{\ast }>=H^{ikjl}S(\mathbf{k}_{1})\delta (\mathbf{k}_{1}-\mathbf{%
k}_{2}) \\
&<&g^{ikm}(\mathbf{k}_{1}\mathbf{)(}g^{jnl}(\mathbf{k}_{2}\mathbf{))}^{\ast
}>=G_{jnl}^{ikm}S(\mathbf{k}_{1})\delta (\mathbf{k}_{1}-\mathbf{k}_{2})
\end{eqnarray*}
where $S(\mathbf{k})$ is a scalar spectral density of the structural
inhomogeneities, $H^{ikjl}$and $G_{jnl}^{ikm}$ are isotropic tensors with a
structure reflecting properties of coefficients $h_{\sigma }^{ik}$ and $%
g^{ikm}$.

Calculations of the non-coherent contributions in the cross-section requires
averaging of the expressions like 
\begin{equation*}
<<W_{\sigma }(\mathbf{k}_{1})W_{\sigma }^{\ast }(\mathbf{k}_{2})\tilde{h}%
_{\sigma }^{ik}(\mathbf{K}-\mathbf{k}_{1})(\tilde{h}_{\sigma }^{jl}(\mathbf{K%
}-\mathbf{k}_{2}))^{\ast }>>
\end{equation*}
. Doing so we use the simplest decoupling of the correlators of this from as
follows: 
\begin{eqnarray*}
&<&<W_{\sigma }(\mathbf{k}_{1})W_{\sigma }^{\ast }(\mathbf{k}_{2})\tilde{h}%
_{\sigma }^{ik}(\mathbf{K}-\mathbf{k}_{1})(\tilde{h}_{\sigma }^{jl}(\mathbf{K%
}-\mathbf{k}_{2}))^{\ast }>>\simeq \\
&<&<W_{\sigma }(\mathbf{k}_{1})W_{\sigma }^{\ast }(\mathbf{k}_{2})>><\tilde{h%
}_{\sigma }^{ik}(\mathbf{K}-\mathbf{k}_{1})(\tilde{h}_{\sigma }^{jl}(\mathbf{%
K}-\mathbf{k}_{2}))^{\ast }>
\end{eqnarray*}
This approximation is justified in the case of sufficienlty weak
inhomogeneities, which is assumed throughout the paper. Within this
approximation the non-coherent cross-section can again be separated into two
terms $I_{W}^{ncoh}$ and $I_{E}^{ncoh}$ with quantities $I$ having the
meaning similar to that in Eq.(\ref{crosscoh}). It is natural to assume
isotropy of the correlation tensors $H^{ikjl}$ and $G_{jnl}^{ikm}$ with
additional symmetry properties following from the symmetries of the initial
mechano-optical and electro-optical coefficients. For the vibrational
contribution $I_{W}^{ncoh}$ in the non coherent scattering one then has 
\begin{equation}
I_{W}^{ncoh}=[H_{1}+H_{2}(\mathbf{e}^{s}\mathbf{e}^{in})^{2}]\sum_{\sigma
}\int d^{3}q\mathrm{Im}\left[ \eta _{\sigma }^{W}(\Omega ,\mathbf{q)}\right]
S(\mathbf{K}-\mathbf{q),}  \label{W-noncoh}
\end{equation}
where $H_{1}$ and $H_{2}$\ are the only independent elements of the tensor $%
H^{ikjl}$. An expression for the non-coherent electromagnetic contribution
is somewhat more cumbersome because of more complicated tensor structure of $%
G_{jnl}^{ikm}$: 
\begin{multline}
I_{E}^{ncoh}=\left[ G_{1}+G_{2}(\mathbf{e}^{s}\mathbf{e}^{in})^{2}\right]
\int d^{3}q\mathrm{Im}\left[ \eta _{ii}^{E}(\Omega ,\mathbf{q}\right] S(%
\mathbf{K}-\mathbf{q)+} \\[0.01in]
\left[ G_{3}\left( e_{i}^{in}e_{j}^{in}+e_{i}^{s}e_{j}^{s}\right)
+G_{4}\left( e_{i}^{s}e_{j}^{in}+e_{i}^{in}e_{j}^{s}\right) \right] \int
d^{3}q\mathrm{Im}\left[ \eta _{ij}^{E}(\Omega ,\mathbf{q)}\right] S(\mathbf{K%
}-\mathbf{q).}  \label{E-noncoh}
\end{multline}

\section{Linear-response functions for disorder-induced polaritons}

\ \ This section of the paper concerns with linear response functions $\eta
_{\sigma }^{W}(\Omega ,\mathbf{q)}$ and $\eta _{ij}^{E}(\Omega ,\mathbf{q)}$%
, which determine Raman scattering cross-section in our system. We start
from dynamical equations describing ionic vibrations in the following form 
\begin{equation}
m_{\alpha }\frac{\partial ^{2}U_{\alpha }^{i}}{\partial t^{2}}+\sum_{\beta ,%
\mathbf{r}_{1}}D_{\alpha \beta }^{ij}(\mathbf{r,r}_{1})U_{\beta }^{j}(%
\mathbf{r}_{1})=e_{\alpha }E^{i}(\mathbf{r).}  \label{dynamics}
\end{equation}
This equation must be complemented by Maxwell's equations 
\begin{equation}
\frac{\partial ^{2}\mathbf{E}}{\partial t^{2}}+c^{2}\mathbf{\nabla \times
\nabla \times E}=-4\pi \frac{\partial ^{2}\mathbf{P}}{\partial t^{2}}
\label{maxwell}
\end{equation}
with polarization density $\mathbf{P}$ defined as 
\begin{equation}
\mathbf{P=}\frac{1}{V}\sum_{\alpha }e_{\alpha }\mathbf{U}_{\alpha }+\chi
_{el}\mathbf{E}.  \label{polarization}
\end{equation}
Indices $\alpha $,$\beta $ label ions within one elementary cell, $U^{i}$ is
a Cartesian component of a displacement vector, $D_{\alpha \beta }^{ij}(%
\mathbf{r,r}_{1})$ is a dynamical matrix describing interaction between ions
of masses $m_{\alpha }$ and charges $e_{\alpha }$. For given radius-vectors $%
\mathbf{r,r}_{1}$ the dynamical matrix can be diagonalized with respect to
sub-cell and Cartesian indices. If $c_{\alpha \sigma }^{i}(\mathbf{r)}$ is a
matrix that diagonalizes the dynamical matrix, the equation of motion (\ref
{dynamics}) and the equation for polarization (\ref{polarization}) can be
rewritten in the following form 
\begin{gather}
\frac{\partial ^{2}W_{\sigma }}{\partial t^{2}}+\sum_{\mathbf{r}%
_{1}}D^{\sigma }(\mathbf{r,r}_{1})W_{\sigma }(\mathbf{r}_{1})=Z_{\sigma }(%
\mathbf{r})d_{\sigma }^{i}E^{i}(\mathbf{r)}  \label{normalmodesequation} \\
\mathbf{P=}\frac{1}{V}\sum_{\sigma }Z_{\sigma }(\mathbf{r})\mathbf{d}%
_{\sigma }\mathbf{W}_{\sigma }+\chi _{el}\mathbf{E},  \notag
\end{gather}
where normal mode $W_{\sigma }(\mathbf{r})$ and its effective charge $%
Z_{\sigma }(\mathbf{r})$ are determined according to 
\begin{eqnarray}
U_{\alpha }^{i}(\mathbf{r}) &=&\sum_{\alpha }c_{\alpha \sigma }^{i}(\mathbf{%
r)}\frac{e_{\alpha }}{\sqrt{m_{\alpha }}}W_{\sigma }(\mathbf{r)}
\label{effectivecharge} \\
Z_{\sigma }(\mathbf{r})d_{\sigma }^{i} &=&\sum_{\alpha }c_{\alpha \sigma
}^{i}(\mathbf{r)}\frac{e_{\alpha }}{\sqrt{m_{\alpha }}}.  \notag
\end{eqnarray}
The vector $d_{\sigma }^{i}$\ is a polarization vector of $\sigma $-th mode.
In a crystal with cubic symmetry all modes can be classified as longitudinal
or transverse ones, making the direction of the polarization vector
independent of the orientation of the elementary cell. Therefore, we can
assume that local inhomogeneities, entering Eq.(\ref{effectivecharge})
through the $\mathbf{r}$-dependence of the diagonalizing matrix $c_{\alpha
\sigma }^{i}(\mathbf{r)}$, result in fluctuations of the effective charge \ $%
Z_{\sigma }(\mathbf{r})$, leaving the directions of the polarization vectors
without change. According to our basic assumption all triply degenerate
modes contributing to the scattering have even parity, so the mean value of
the effective charge is equal to zero. Correlation properties of the
inhomogeneous effective charge are specified by means of the autocorrelation
function $K(r)=\langle Z_{\sigma }(\mathbf{r})Z_{\sigma }(\mathbf{0})\rangle 
$ or its spectral density $S(k)=\int d^{3}rK(r)\exp (i\mathbf{kr)}$. In what
follows we assume that the correlation function, $K(r)$, and spectral
density $S(k)$ are given as 
\begin{eqnarray}
K(r) &=&\langle Z_{\sigma }^{2}\rangle \exp (-rk_{c})
\label{correlation_function} \\
S(k) &=&\langle Z_{\sigma }^{2}\rangle \frac{k_{c}}{\pi ^{2}}\frac{1}{%
(k^{2}+k_{c}^{2})^{2}},  \notag
\end{eqnarray}
where $\langle Z_{\sigma }^{2}\rangle $ is an rms fluctuation of the
effective charge, and $k_{c}$\ is a correlation wave number, which is the
inverse of a correlation radius and set the spatial scale of the
inhomogeneities. We will also assume that the diagonalized dynamic matrix $%
D^{\sigma }(\mathbf{r,r}_{1})$ is nonrandom, so that the mechanical
properties of the system are not affected by the disorder.\ 

\subsection{Vibrational response function}

Linear response functions $\eta _{\sigma }^{W}(\Omega ,\mathbf{q)}$ and $%
\eta _{ij}^{E}(\Omega ,\mathbf{q)}$can be found from the solutions of Eq.(%
\ref{normalmodesequation}) and Eq.(\ref{maxwell}) with appropriate source
terms on the right hand side of the equations. We will start by determining
the mechanical response function $\eta _{\sigma }^{W}(\Omega ,\mathbf{q)}$%
\textbf{. }For this purpose we\ add an external force $\mathbf{F}$ to the
right hand side of the equation Eq.(\ref{normalmodesequation}), and our aim
is to find an equation for the averaged over the disorder amplitude, $%
\langle W_{\sigma }(\Omega ,\mathbf{K}\rangle $ . Carrying out the
calculations with the use of a perturbation approach with respect to the rms
fluctuation of the effective charge, $d^{2}$, discussed in detail, for
instance, in Ref.\cite{susceptibilities}, we can find with the accuracy to
the first nontrivial correction (Bourret approximation) 
\begin{multline}
\left( \Omega ^{2}-\varepsilon _{K}^{2}\right) \langle W_{\sigma }(\Omega ,%
\mathbf{K}\rangle - \\
\frac{4\pi }{\epsilon _{\infty }V}\Omega ^{2}\sum_{\alpha }\langle W_{\alpha
}(\Omega ,\mathbf{K}\rangle \int d^{3}qS(\mathbf{K}-\mathbf{q})d_{\sigma
}^{i}(\mathbf{K}-\mathbf{q)}d_{\alpha }^{j}(\mathbf{K}-\mathbf{q)}%
G_{em}^{ij}(\mathbf{q)=}F_{K}.  \label{averaged_displac}
\end{multline}
We introduce here an initial dispersion law of the vibrations, $\varepsilon
_{K}^{2}$, describing triply degenerate (in the absence of interaction with
electromagnetic field) mode. In the long wavelength limit the dispersion law
assumes the form $\varepsilon _{K}^{2}=\omega _{0}^{2}+v^{2}K^{2}$. Tensor $%
G_{em}^{ij}(\mathbf{q)}$ in Eq.(\ref{averaged_displac}) is the Green's
function of Maxwell's equation (\ref{maxwell}) in the absence of the
coupling with the vibrations 
\begin{equation*}
G_{em}^{ij}(\mathbf{q)=}\frac{1}{\Omega ^{2}-c^{2}q^{2}}(\delta _{ij}-\frac{%
q_{i}q_{j}}{q^{2}})+\frac{1}{\Omega ^{2}}\frac{q_{i}q_{j}}{q^{2}}.
\end{equation*}
High-frequency dielectric constant $\epsilon _{\infty }$ takes into account
the electronic contribution into the polarization, and speed of light $c$ in
the previous equation is corrected for $\epsilon _{\infty }$. Nondiagonal
terms (with respect to the polarization indices $\alpha $, $\sigma $) in Eq.(%
\ref{averaged_displac}) can be shown to vanish after the integration over $%
\mathbf{q}$, so we only have to evaluate $\eta _{L}^{W}$ for the
longitudinal mode and $\eta _{T}^{W}$ for transverse modes. Evaluating the
integral in Eq.(\ref{averaged_displac}) one has to take into account that
typical values of the correlation wave number $k_{c}$ determining the scale
of the inhomogeneities can be of the order of magntidue between$%
10^{6}cm^{-1} $ and $10^{8}cm^{-1}$ while the value of polariton wave
numbers is of the order of $10^{3}cm^{-1}$ or $10^{4}cm^{-1}$. Another
important observation is that the frequency $\omega _{c}=ck_{c}$
corresponding to the correlation wave number is much greater than all other
characteristic frequencies of the system. Therefore, one can neglect any
effects of retardation, which are of the order of $(\Omega /$ $\omega
_{c})^{2}$, and evaluating all the remaining integrals with the accuracy to $%
(K/k_{c})^{2}$, obtain longitudinal and transverse vibrational response
functions in the following form 
\begin{equation}
\eta _{L,T}^{W}(\Omega ,K)=\frac{1}{\Omega ^{2}-\varepsilon
_{L,T}^{2}(K)-2i\Omega \Gamma _{in}},  \label{W-response}
\end{equation}
where 
\begin{equation}
\varepsilon _{L}^{2}=\omega _{0}^{2}+\Lambda ^{2}+(v^{2}-\frac{1}{6}\frac{%
\Lambda ^{2}}{k_{c}^{2}})K^{2},\varepsilon _{T}^{2}=\omega _{0}^{2}+(v^{2}+%
\frac{1}{6}\frac{\Lambda ^{2}}{k_{c}^{2}})K^{2},  \label{new_vibr_dispesrion}
\end{equation}
The parameter $\Gamma _{in}$ takes into account initial damping of the
vibrations, which is not related to the interaction with the field and is
determined by intrinsic relaxation processes such as, for instance,
anharmonism. This parameter can also include some effects of an additional
scattering due to inhomogeneities that are intrinsic to the ion system, for
example, spatial fluctuations of the dynamical matrix. Parameter $\Lambda $
introduced above is an effective coupling parameter defined as 
\begin{equation*}
\Lambda ^{2}=\frac{4\pi \langle Z_{\sigma }^{2}\rangle }{V\epsilon _{\infty }%
}.
\end{equation*}

It can be seen from these expressions that the vibrational response function
does not contain effects of the disorder-induced interaction between
vibrations and transverse electromagnetic field. This happens because the
effective relaxation parameter $ck_{c}$, which is responsible for the
scattering of the phonons into electromagnetic waves, is much greater than
the effective coupling parameter $\Lambda $. Therefore, according to the
general results obtained in Ref.\cite{DICR}, the coupling is ineffective and
phonon properties are practically unaffected by the interaction.

However, the interaction between phonons and a longitudinal macroscopic
field modifies their properties. In particular, even zero-mean random
effective charge lifts the degeneracy between longitudinal and transverse
phonon modes giving rise to the same $LT$-splitting that occurs in regular
ideal crystals. It is more interesting that disorder also modifies the
parameter $v $, which is responsible for the sign and strength of the
spatial dispersion of the modes. The correction $\frac{1}{6}\frac{\Lambda
^{2}}{k_{c}^{2}}$ has opposite signs for longitudinal and transverse modes,
and its order of magnitude can be comparable to the initial parameter $v$.
In ideal crystals the longitudinal electrostatic field is uniform and exerts
an equal force upon ions at different sites, resulting in the uniform lift
of the longitudinal mode. In the presence of disorder, however, the field is
not uniform anymore giving rise to certain corrections to the inter-ion
potential. These corrections finally manifest themselves as modification of
the parameter of the spatial dispersion $v$, and can be of the same order of
magnitude as the initial potential causing substantial modification of the
initial spectrum.

\subsection{Electromagnetic response function.}

In order to evaluate the electromagnetic response function $\eta
_{ij}^{E}(\Omega ,K)$ one has to insert a source term into the Maxwell
equation (\ref{maxwell}). According to the fundamentals of the linear
response theory this source term should have the form of an external
polarization \cite{Loudon}. Taking the source term into account and
averaging the equations over the disorder in the same manner as above we
finally arrive at the following equation for the average amplitude of the
electromagnetic field 
\begin{equation}
\left[ \left( \Omega ^{2}-c^{2}K^{2}-\frac{4\pi }{\epsilon _{\infty }V}%
\Omega ^{2}\int d^{3}q\frac{S(\mathbf{K}-\mathbf{q})}{\Omega
^{2}-\varepsilon _{q}^{2}}\right) \delta _{ij}+c^{2}K_{i}K_{j}\right]
\langle E_{j}(\Omega ,\mathbf{K}\rangle \mathbf{=}\frac{4\pi \Omega ^{2}}{%
\epsilon _{\infty }}.P_{ext}^{i}.  \label{averaged_Electric}
\end{equation}
The electromagnetic response function $\eta _{ij}^{E}(\Omega ,K)$ is usually
determined as \cite{Loudon} 
\begin{equation}
\eta _{ij}^{E}(\Omega ,K)=\frac{4\pi \Omega ^{2}}{\epsilon _{\infty }V}\beta
_{ij}(\Omega ,K),  \label{E-response}
\end{equation}
where $\beta _{ij}(\Omega ,K)$ is the inverse matrix for the left-hand side
matrix of Eq.\ref{averaged_Electric}and is given as usual by the sum of the
transverse and longitudinal parts 
\begin{eqnarray}
\beta _{ij}(\Omega ,K) &=&\beta _{T}(\delta _{ij}-\hat{K}_{i}\hat{K}%
_{j})+\beta _{L}\hat{K}_{i}\hat{K}_{j},  \label{beta-tensor} \\
\beta _{T} &=&\frac{1}{\left( \Omega ^{2}-c^{2}K^{2}-\frac{4\pi }{\epsilon
_{\infty }V}\Omega ^{2}\int d^{3}q\frac{S(\mathbf{K}-\mathbf{q})}{\Omega
^{2}-\varepsilon _{q}^{2}}\right) }  \notag \\
\beta _{L} &=&\left( \beta _{T}^{-1}+c^{2}K^{2}\right) ^{-1},  \notag
\end{eqnarray}
where $\mathbf{\hat{K}}$ is a unit vector in the direction of $\mathbf{K.}$%
The integral with the spectral density given by Eq.(\ref
{correlation_function}) can be easily calculated to yield the following
expression for $\beta _{T}$%
\begin{equation}
\beta _{T}=\frac{\Omega ^{2}-\varepsilon _{K}^{2}-\varkappa ^{2}(\Omega
)-2i\Omega (\Gamma _{in}+\Gamma _{s})}{(\Omega ^{2}-c^{2}K^{2})\left( \Omega
^{2}-\varepsilon _{K}^{2}-\varkappa ^{2}(\Omega )-2i\Omega (\Gamma
_{in}+\Gamma _{s})\right) -\Lambda ^{2}\Omega ^{2}},  \label{beta_transv}
\end{equation}
where 
\begin{eqnarray}
\varkappa ^{2}(\Omega ) &=&v^{2}k_{c}^{2}-2vk_{c}\Omega \left[ \left( 1-%
\frac{\omega _{0}^{2}}{\Omega ^{2}}\right) ^{2}+\frac{4\Gamma _{in}^{2}}{%
\Omega ^{2}}\right] ^{1/4}\sin \left[ \frac{1}{2}\arctan \left( \frac{%
2\Omega \Gamma _{in}}{\Omega ^{2}-\omega _{0}^{2}}\right) \right]  \notag \\
\Gamma _{s}(\Omega ) &=&vk_{c}\left[ \left( 1-\frac{\omega _{0}^{2}}{\Omega
^{2}}\right) ^{2}+\frac{4\Gamma _{in}^{2}}{\Omega ^{2}}\right] ^{1/4}\cos %
\left[ \frac{1}{2}\arctan \left( \frac{2\Omega \Gamma _{in}}{\Omega
^{2}-\omega _{0}^{2}}\right) \right] ,  \label{gamma_s}
\end{eqnarray}
and we again introduced an initial phonon damping $\Gamma _{in}$. The poles
of the function $\beta _{T}$ corresponds to the dispersion law of the
transverse electromagnetic waves modified due to the random interaction with
the phonons. The equation for the poles 
\begin{equation}
(\Omega ^{2}-c^{2}K^{2})\left( \Omega ^{2}-\varepsilon _{K}^{2}-\varkappa
^{2}(\Omega )-2i\Omega (\Gamma _{in}+\Gamma _{s})\right) -\Lambda ^{2}\Omega
^{2}=0  \label{poles}
\end{equation}
resembles, as it was pointed out in Ref.\cite{DICR}, an equation for eigen
frequencies of two waves propagating in a deterministic ''effective''
medium. The functions $\varkappa ^{2}(\Omega )$ and $\Gamma _{s}(\Omega )$
describe a renormalized dispersion law of phonons in this medium and their
damping respectively. Actually, this equation describes the interaction of
the coherent (average) electromagnetic wave with non coherent (scattered )
ion vibrations, and its solutions give a dispersion law for the coherent
component of the electromagnetic waves modified by the disordered coupling
with the phonons. An analysis of the simplified version of the Eq.(\ref
{poles}) carried out in Ref.\cite{DICR,DIP}) showed that this equation can
yield a solution with two well-defined branches of ''disorder-induced''
polaritons if $\Lambda >\Lambda _{cr}$, where $\Lambda _{cr}$ was estimated
in Ref.\cite{DIP} as $\Lambda _{cr}\simeq v^{2}k_{c}^{2}/\omega _{0}$ in the
model without the initial relaxation $\Gamma _{in}$. However, because of the
small value of the parameter $v$ of the phonon spatial dispersion, the
initial phonon damping cannot be actually neglected, and a more realistic
expression for $\Lambda _{cr}$ can be shown to be $\Lambda _{cr}\simeq \sqrt{%
(\Gamma _{in}/\Omega )}\left( vk_{c}+\sqrt{\Omega \Gamma _{in}}\right) $.
Numerically the order of magnitude of $\Lambda _{cr}$ does not differ from
the estimate presented in Ref.\cite{DIP}, so the optimistic conclusion of
that paper regarding the possibility of observing disorder-induced
polaritons in real systems still holds.

Two branches of the disorder-induced polaritons can manifest themselves as
two maxima on the frequency dependence of the transverse component of
electromagnetic response function $\eta _{ij}^{E}(\Omega ,K)$. Detailed
study of the relationship between maxima of response functions and solutions
of the dispersion equations was carried out in Ref.\cite{susceptibilities}
in the case of the disorder-induced magnetoelastic resonance. Applying the
general results obtained in Ref.\cite{susceptibilities} to our situation we
can conclude that two maxima \ appear on $\eta _{ij}^{E}(\Omega ,K)$\ when a
gap between two solutions of the dispersion equation (\ref{poles}) at the
crossing point becomes greater then $\Lambda _{cr}$. Since $\Lambda _{cr}$
is the inversed decay time for the solutions at the resonance, the stated
condition for two maxima to appear has a clear physical meaning - the
frequency interval between two peaks must be greater than their width.

Though qualitatively the conclusions drawn from the simplified model of Ref. 
\cite{susceptibilities} remain valid for our more complicated situation,
there is an important difference. The effective damping parameter in our
case is frequency dependent. This fact leads to an asymmetrical form of the
frequency dependence of the respective response function. From Eq.(\ref
{gamma_s}) one can see that parameter $\Gamma _{s}$ decreases with the
increase of frequency. One can expect, therefore, that the low frequency
maximum of the responce function will be lower and wider than the maximum
corresponding to the upper polariton branch. The plots presented in the
subsequent section of the paper support this conclusion.

The fact that vibrational and electromagnetic response functions demonstrate
such a different behavior is typical for systems with disorder induced
crossing resonances. The simple qualitative interpretation of this fact was
given in Ref.\cite{susceptibilities}, where the concept of two effective
media was introduced. According to this concept, averaging over the disorder
results in two different effective media for each of the interacting waves.
The most important difference between the media is in their relaxation
properties. The order of magnitude of the effective relaxation parameters is
determined, among other things, by the slope of the initial dispersion
curves and the correlation radius of the disorder. The former in the case of
electromagnetic waves is just speed of light and is much greater than the
corresponding parameter $v$ for phonons. As a result, a dispersion law for
coherent phonons is determined by the interaction with electromagnetic waves
propagating in strongly absorbing effective media; the modification of the
initial dispersion law is negligibly small in this situation. As for
coherent electromagnetic waves, their dispersion law is modified due to the
coupling with phonons in the effective media with rather weak absorption.
The coupling is much more effective in this situation. Therefore, the
initial dispersion curve splits at the resonance resulting in two maxima on
the electromagnetic response function.

\section{Frequency dependence of the scattering cross section}

The response functions presented in the previous section can be used to
study Raman scattering cross- section in our system. The results for the
coherent part of the cross-section are obtained by a substitution of
imaginary parts of Eq.(\ref{W-response}) into Eq.(\ref{Wcoh}) for the ionic
contribution into the coherent cross-section, and the imaginary part of Eq.(%
\ref{E-response}) into Eq.(\ref{Ecoh}) for the electromagnetic contribution.
As a result, the ionic contribution is 
\begin{multline}
I_{W}^{coh}=2h^{2}|\epsilon _{ikm}||\epsilon
_{jlp}|e_{i}^{s}e_{j}^{s}e_{k}^{in}e_{l}^{in}\Omega \Gamma _{in}\times \\
\left\{ \frac{\hat{K}_{m}\hat{K}_{p}}{\left[ \Omega ^{2}-\varepsilon
_{L}^{2}(K)\right] ^{2}+4\Omega ^{2}\Gamma _{in}^{2}}+\frac{2\left( \delta
_{mp}-\hat{K}_{m}\hat{K}_{p}\right) }{\left[ \Omega ^{2}-\varepsilon
_{T}^{2}(K)\right] ^{2}+4\Omega ^{2}\Gamma _{in}^{2}}\right\}  \label{IWcoh}
\end{multline}
and electromagnetic part of the cross-section has the following form 
\begin{multline}
I_{E}^{coh}=g^{2}|\epsilon _{ikm}||\epsilon
_{jlp}|e_{i}^{s}e_{j}^{s}e_{k}^{in}e_{l}^{in}\frac{8\pi \Omega \Lambda
^{2}(\Gamma _{in}+\Gamma _{s})}{V\epsilon _{\infty }}\times \\
\left[ 2\left( \delta _{mp}-\hat{K}_{m}\hat{K}_{p}\right) \mathrm{Im}\beta
_{T}^{\prime }(\Omega ,K)+\hat{K}_{m}\hat{K}_{p}\mathrm{Im}\beta
_{L}^{\prime }(\Omega ,K)\right] .  \label{IEcoh}
\end{multline}
The $\mathrm{Im}\beta _{T}^{^{/}}(\Omega ,K)$ term in Eq.(\ref{IEcoh})
represents the contribution due to scattering from transverse modes 
\begin{equation}
\mathrm{Im}\beta _{T}^{\prime }=\frac{2\Omega ^{4}}{\left[ (\Omega
^{2}-c^{2}K^{2})(\Omega ^{2}-\varepsilon _{K}^{2})-\Lambda ^{2}\Omega ^{2}%
\right] ^{2}+4\Omega ^{2}(\Omega ^{2}-c^{2}K^{2})^{2}(\Gamma _{in}+\Gamma
_{s})^{2}},  \label{IEcohT}
\end{equation}
whereas the second term corresponds to the scattering due to the
longitudinal mode 
\begin{equation}
\mathrm{Im}\beta _{L}^{\prime }=\frac{1}{(\Omega ^{2}-\varepsilon
_{K}^{2}-\Lambda ^{2})^{2}+4\Omega ^{2}(\Gamma _{in}+\Gamma _{s})^{2}},
\label{IEcohL}
\end{equation}
where we neglect the term $\varkappa ^{2}(\Omega )$ defined in Eq.(\ref
{gamma_s}), which can be shown to be small.

The polarization dependence of the coherent scattered intensity is mostly
determined by the symmetry of the underlying structure, therefore we shall
be focused upon a frequency profile of the intensity, which has more
universal significance. We would like to note, however, that the suggested
average symmetry of the system results in vanishing of both coherent
contributions when the polarization of the scattered wave coincides with the
polarization of the incident wave. This feature of the considered structure
can be used to separate the coherent contribution from the non coherent one.
In what follows we restrict our consideration to the perpendicular
polarization geometry. In this case the angular dependent factors in Eqs.(%
\ref{IEcoh},\ref{IWcoh}) take the form $1-\hat{K}_{in}^{2}$ for the
transverse component , and $\hat{K}_{in}^{2}$ for the longitudinal one,
where $\hat{K}_{in}$ is a component of \ $\mathbf{\hat{K}}$ \ in the
direction of propagation of the incident wave, $\hat{K}_{in}\approx \sin
(\theta /2)$, where $\theta $ is an angle between wave vectors of the
incident and scattered waves.

Polariton effects manifest themselves in Raman scattering only at small
scattering angles, corresponding to small wave numbers $K$. The main
contribution into the coherent scattering in this situation comes from the
transverse components, which depends upon the scattering angle as $\cos
^{2}(\theta /2)$, while the contributions from the longitudinal components
is small as $\sin ^{2}(\theta /2)$. The frequency profile of the spectrum is
determined by the sum of two parts of the scattering cross-section: ionic
contribution Eq.(\ref{IWcoh}) and electromagnetic contribution Eq.(\ref
{IEcoh}). The former component of the cross-section does not contain any
traces of polariton effects as was explained in the previous sections of the
paper. Its frequency dependence has a simple form with one maximum, which is
situated at the frequency of TO phonons. Due to weak phonon dispersion,
position of this maximum almost independent of scattering angle $\theta $.
Effects of disorder-induced polaritons can appear in the scattering spectrum
only due to the electromagnetic contribution to the cross-section. If the
system under study allows for disordered polaritons to arise, the transverse
response function $\mathrm{Im}\beta _{T}$ has two peaks at the polariton
frequencies, which are shifted with respect to the TO frequency. The
resulting coherent intensity has, in this situation, three peaks: one in \
the center due to the ionic contribution, and two others due to the
disorder-induced polaritons (see Fig.1, where electro-optical and mechanical
coherent intensity are plotted together. A small bump on the electro-optical
spectrum corresponds to a small longitudinal contribution at LO frequency.)
Positions and hight of polariton peaks significantly depend upon polariton
wave number $\mathbf{K}$, which in turn is determined by the scattering
angle. These dependence can be obtained on the bases of the general analysis of susceptibilities presented in Ref.\cite{susceptibilities}. At $\mathbf{K}$ much smaller than the resonance value, one can
observe only one wide peak of $\mathrm{Im}\beta _{T}$ at LO frequency. The total
coherent spectrum in this case will consist of two maxima corresponding to
LO and TO phonons. Upon increasing $\mathbf{K}$ the original LO maximum will
start shifting toward high frequencies while becoming narrower and taller. Simultaneously an additional
maximum will emerge at
frequencies below the frequency of TO phonons. The overall picture in resonance region,
hence, will contain three maxima with the central one corresponding to TO
phonons, and two others resulting from disorder-induced polariton
excitations. According to the general results of the theory of disorder-induced crossing resonance \cite{DICR} the width of the polariton peaks allows for tracking the dependence between location of the peak and polariton wave number $\mathbf{K}$, i.e. the dispersion law of the disorder-induced polaritons. If the crossing resonance does take place one would observe in this case the parts of two polariton branches described in Ref.\cite{DICR,DIP}. Increasing $\mathbf{K}$ beyond the resonance value will move the
higher frequency maximum toward even higher frequency while making it
sharper. The low frequency maximum at the same time will be approaching TO
frequency and broadening. Finally the upper polariton branch will
become unvisible, and the lower one will be indistinguishable from TO
phonons. At the same time the contribution from longitudinal cross-section
will rise, therefore one will again observe the spectrum with two maxima at
LO and TO frequencies.

If the effective coupling parameter $\Lambda $ is too small for the
disorder-induced crossing resonance to occur, then all one could observe is
a two maximum spectrum whose shape only weakly depends upon the scattering
angle.

Now let us consider non coherent scattering. Substituting Eq.(\ref
{W-response}) for the vibrational response function in Eq.(\ref{W-noncoh})
for the non coherent cross-section, and evaluating the integral with the
spectral density given by Eq.(\ref{correlation_function}), one arrives at
the following expression for $I_{W}^{ncoh}$%
\begin{multline}
I_{W}^{ncoh}=2[H_{1}+H_{2}(\mathbf{e}^{s}\mathbf{e}^{in})^{2}]\Omega \times
\\
\left\{ \frac{\Gamma _{L}^{eff}(\Omega )}{\left( \Omega ^{2}-\omega
_{L}^{2}\right) ^{2}+4\Omega ^{2}\left[ \Gamma _{L}^{eff}(\Omega )\right]
^{2}}+\frac{\Gamma _{T}^{eff}(\Omega )}{\left( \Omega ^{2}-\omega
_{T}^{2}\right) ^{2}+4\Omega ^{2}\left[ \Gamma _{L}^{eff}(\Omega )\right]
^{2}}\right\} ,  \label{IW_nonch}
\end{multline}
where $\omega _{T}^{2}=\omega _{0}^{2}+v_{T}^{2}k_{c}^{2}$, $\omega
_{L}^{2}=\omega _{0}^{2}+v_{L}^{2}k_{c}^{2}+\Lambda ^{2}$ and 
\begin{multline}
\Gamma _{L,T}^{eff}(\Omega )=\Gamma _{in}+v_{L,T}k_{c}\left[ \left( 1-\frac{%
\omega _{L,T}^{2}}{\Omega ^{2}}\right) ^{2}+\frac{4\Gamma _{in}^{2}}{\Omega
^{2}}\right] ^{1/4}\times  \label{Gamma_eff} \\[0.02in]
\cos \left[ \frac{1}{2}\arctan \left( \frac{2\Omega \Gamma _{in}}{\Omega
^{2}-\omega _{L,T}^{2}}\right) \right] ,
\end{multline}
and $v_{T,L}^{2}=(v^{2}\pm \frac{1}{6}\Lambda ^{2}/k_{c}^{2})$ are the
phonon dispersion parameters modified due to the interaction with the
electromagnetic field (see Eq.(\ref{new_vibr_dispesrion})). we have
neglected here $K$-dependence of the phonon dispersion laws. Eq.(\ref
{IW_nonch}) has a structure similar to the expression for the coherent
vibrational intensity except for its independence from the scattering angle.
Therefore, transverse and longitudinal components have equal weights. The
positions and the widths of the both peaks, however, differ from the
coherent scattering. Fluctuations of the mechano-optical coefficients result
in a shift of the transverse and longitudinal frequencies by $%
v_{T,L}^{2}k_{c}^{2}$ respectively, and cause an increase in the width given
by the second term in Eq.(\ref{Gamma_eff}). This expression is similar to
Eq.(\ref{gamma_s}), which determines the widths of the maxima of the
coherent electromagnetic contribution, with the obvious replacement of the
parameters.

The analysis of the non-coherent electromagnetic scattering is somewhat more
cumbersome because of the more complicated polarization dependence. We would
like to point out two simplifying details. First, a contribution from the
transverse part of the electromagnetic response function has the order of $%
(\Omega /ck_{c})^{3}$ and can be neglected. Second, we can neglect $K$%
-dependence in the corresponding integrals. As a result, the expression for
non coherent electromagnetic part of the scattering intensity takes the
following form 
\begin{equation}
I_{E}^{ncoh}=\left[ \tilde{G}_{1}+\tilde{G}_{2}(\mathbf{e}^{s}\mathbf{e}%
^{in})^{2}\right] \frac{4\pi \Lambda ^{2}}{V\epsilon _{\infty }}\frac{\Omega
(\Gamma _{in}+\Gamma _{s})}{\left( \Omega ^{2}-\omega _{0}^{2}-\Lambda
^{2}\right) ^{2}+4\Omega ^{2}(\Gamma _{in}+\Gamma _{s})^{2}},
\label{IE_noncoh}
\end{equation}
where $\tilde{G}_{1}$ and $\tilde{G}_{2}$ are certain combinations of the
original coefficients $G$. One can see that the non coherent electromagnetic
contribution is identical to the coherent longitudinal contribution.

A sample plot for the total scattering intensity is shown in Fig.2. In
generating plots in Fig.1 and Fig.2 we choose parameters typical for the
systems under consideration, the most important of which are the coupling
parameter, internal phonon damping rate, electro-optical and mechano-optical
coefficients. The coupling parameter $\Lambda $ was chosen to be an order of
magnitude smaller than the resonance frequency, and phonon damping was
assumed to be two orders of magnitudes smaller than the frequency. The
relative value of the electro-optical and mechano-optical parameters was
chosen to make respective contributions into the scattering intensity be of
the same order of magnitude. One can see that the total scattering retain the three-peak feature of the coherent contribution. The peak at the center is formed as a result of superposition of the mechanical coherent cross-section and all the non coherent contributions. Obviously the width of the non coherent maxima for values of parameters chosen for the plot appears to be to large to reveal different types of non coherent contributions, which formally should have appeared at frequencies shifted with respect to coherent TO and LO frequencies. Since the values of the most parameters required to make more explicit predictions of the shape of the spectra are not known  we cannot discuss details of the spectra.  The figures presented demonstrate nevertheless that under
favorable circumstances non coherent contribution would not mask effects due
to disordered polaritons in Raman scattering intensity. We can conclude, therefore, that one could hope to observe the polariton-related effects in materials whose average structure forbid polariton formation by studing  the spectra for different scattering angles as explained above.

\section{Conclusion}

In this paper we develop a theory of Raman scattering of light due to
disorder-induced polaritons. We consider a material that on average retains
a diamond-like cubic structure. This structure has only one triply
degenerate optic phonon mode of even parity, which is dipole-inactive in an
ideal structure. Local random distortion of the structure, however, gives
rise to a random effective charge of the mode and respectively to an
effective random coupling with electromagnetic field. The distortion can be
induced by different means such as application of pressure \cite{pressure},
using thin films \cite{films}, or amorphization. Among materials with the
structure of diamond, for example, $Si$ and $Ge$ exist in amorphous
versions. Raman scattering in the systems considered in the paper has an
interesting feature, which does not show up in the case of ideal structures.
In crystalline materials with center of symmetry there exists an ''exclusion
rule'' for dipole infra-red activity and Raman activity. Dipole-active
modes, which form polariton states, do not participate in the first-order
Raman scattering. Therefore, in ideal crystals the problem of polariton
Raman scattering arises only in structures without a center of symmetry. In
our situation we deal with a mode which is, on average, dipole inactive, and
is, therefore, Raman active. At the same time this mode is coupled to
electromagnetic field due to fluctuating effective charge, and induces
random electromagnetic field. This field has zero average amplitude, but its
intensity is not zero, and it makes sense to take into account the
modulation of the electron polarizability due to this field. Components of
the respective tensor of the susceptibility derivatives are, of course,
random functions and contribute to the non coherent scattering. The average
symmetry of the structure does not forbid, however, this tensor from having
non-zero \emph{average }value. If this is the case, then one can observe 
\emph{coherent }electromagnetic component of the scattering cross-section
from disorder-induced polaritons. Assuming that the average electro-optical
coefficients in our situation do not differ in magnitude from more standard
situations we show that polariton contribution into Raman scattering can
indeed be observed on the non-coherent background.

Raman spectra of amorphous materials have been studied for many years (see,
for example, the latest paper in Ref\cite{review} and references therein).
Their interpretation, however, is rather complicated, therefore, it is
difficult to determine if the disorder-induced polaritons have been already
observed. An accurate identification of disorder-induced polaritons in the
spectra requires special experiments. We considered the behavior of the
Raman cross-section under the change of the scattering angle, and determined
which features of the spectra are specific for disorder-induced polaritons.
On the basis of our analysis we think that comparison of spectra measured at
different scattering angles could reveal the effects caused by
disorder-induced polaritons. We suggest that crystals with locally distorted
structures rather than amorphous materials are the best candidates for
disorder-induced polariton experiments. Additional  means to interpret
experimental Raman spectra are provided by the possibility to compare
spectra of disordered materials with their crystalline counterparts. In this
case one has to take into account that our model does not explicitly
describe the shifting and broadening of the spectra due to inhomogeneities
intrinsic to the phonon subsystem. We incorporated the broading
phenomenologically by means of the initial damping parameter $\Gamma _{in}$,
and can also assume that initial phonon TO frequency is already corrected
for the shift due to the intrinsic phonon scattering. Such consideration is valid if one assumes
that intrinsic phonon parameters of the system (density and force constants)
fluctuate independently of the phonon-photon coupling parameter. This is a
legitimate assumption since a correlator between, for instance, the force
constants and the coupling parameters should be the tensor of rank three.
Such a tensor must vanish in a system with the average structure displaying
a center of symmetry. Therefore, one can say that the central peak of the
predicted three-peak spectrum corresponds to the broaden and shifted peak,
which would correspond to TO phonon peak in the abscence of the polariton
effects.

Among other results we would like to mention the modification of phonon
dispersion laws due to the fluctuating longitudinal electromagnetic field.
Our calculations show that in random systems the longitudinal field modifies
both the fundamental frequency of phonons and the slope of their dispersion
curves. The last effect means that fluctuating component of longitudinal
field contributes into interactions between atoms from different elementary
cells. It is interesting, that this contribution is found in transverse as
well as in longitudinal modes of the vibrations, but with opposite signs.
The fluctuating field weakens longitudinal components of the interatomic
forces, and enhances the transverse ones.

\section*{Acknowledgments}

This work was supported by the NATO Science program and Cooperation Partner
Linkage Grant No HTECH 960919, and by the NATO Networking Supplement Grant
No 971209.

The authors are grateful to A.A. Maradudin and J.L. Birman for useful
discussions of the paper.

\newpage

\section*{Figure Caption}

\noindent Fig.1 Double peacked line shows electro-optical contribution to
coherent scattering intensity. The single peak line is the coherent
mechanical contribution. The frequency on the plot is normalized by the
resonance frequency. \noindent Fig.2 Total scattering intensity with three
peaks. The left one presents the contribution from the lower polariton
branch, the right one from the upper polariton branch. The peak in the
middle is the resulting contribution from TO phonons and non-coherent
scattering. Upon increase of the non-coherent scattering the left peak will
dissapear first, and yet the remaininmg double-peaked structure will present
a clear indication on the existence of the disordered polartions.

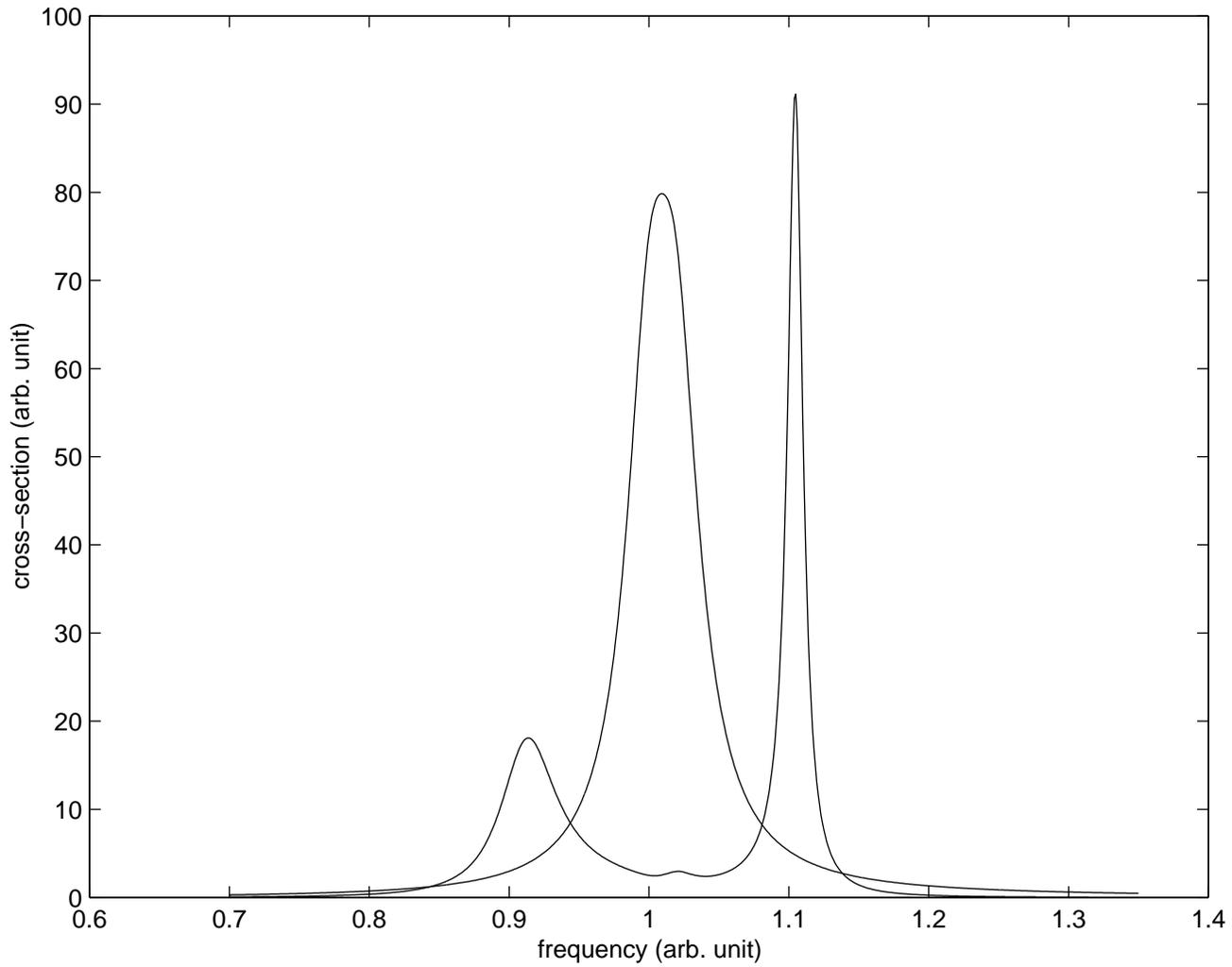
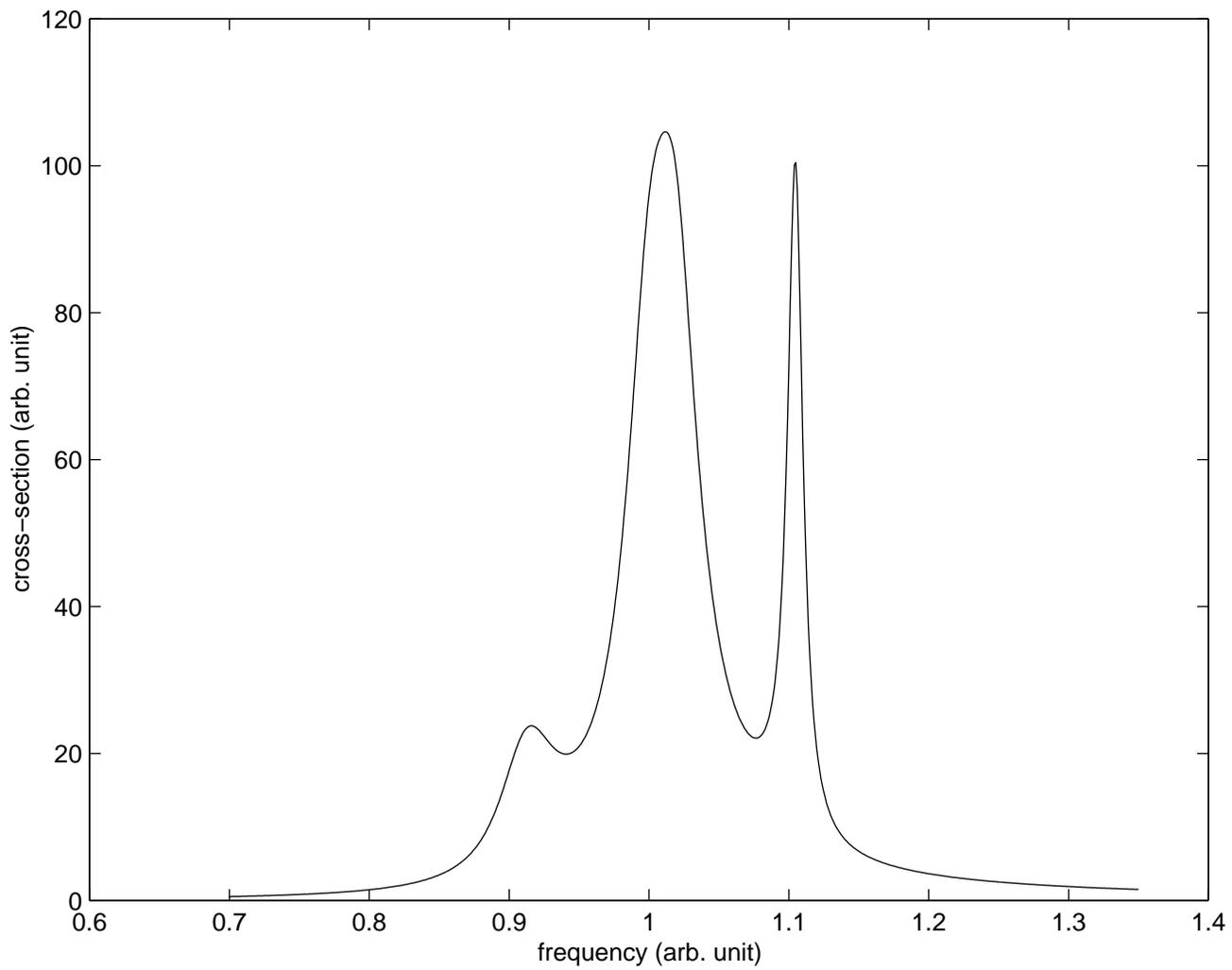
\end{document}